\documentclass[12pt]{article}
\input{amssymb.sty}

\catcode`\@=11
\def\numberbysection{\@addtoreset{equation}{section}
        \def\theequation{\thesection.\arabic{equation}}}
\numberbysection

\newcommand{\be}{\begin{equation}}
\newcommand{\ee}{\end{equation}}
\newcommand{\bea}{\begin{eqnarray}}
\newcommand{\eea}{\end{eqnarray}}

\begin{document}
 
\def\Z{{\mathbb Z}}
\def\N{{\mathbb N}}
\def\R{{\mathbb R}}
\def\Q{{\mathbb Q}}
\def\C{{\mathbb C}}
\def\P{{\mathbb P}}

\def\la{\lambda}
\def\si{\sigma}


\vskip 2truecm

\centerline{{\huge Explicit modular 
formulae       }}

\centerline{{\huge   and symmetries of RCFT's.  I }}

\vskip 2truecm
\centerline{{\large Antoine Coste}}\smallskip
\centerline{{\it CNRS\footnote{ UMR 8627 }
             Laboratory of Theoretical Physics}}
\centerline{{\it building 210, Paris  University  }}
\centerline{{\it 91405 Orsay cedex, France }}
\centerline{{\it coste@math.uni-frankfurt.de   }}

\vskip 1truecm  
     
\centerline{{\it and      }}
\vskip 1truecm 
\centerline{{\it  FB Math. Goethe Univ.  }}
\centerline{{\it R. Mayer Strasse 8      }}
\centerline{{\it D 60054 Frankfurt am Main, Germany     }}

\vskip 1.5truecm

\centerline{{\bf Abstract.}}\vskip .3truecm

We derive compact formulae for  modular transformation 
matrices of Wess-Zumino-Witten (WZW) affine 
characters. 
We  start in this text which 
 is conceived as the first of a series
 with the simple case of  $A_1 $ 
 algebra at positive level
 $k=n-2$, for which we can easily provide 
some description of 
isometry group and  genus formula in a special case. 
 
We also point to 
 general features of these expressions, 
formulating and proving  theorems for RCFT's which seem  
 new.

\vskip 1.cm

\section{Mathematical and Physical origin} 
 
Within the framework of 2d critical phenomena
(so called "CFT's"), 
(our favourite primer is \cite{ID}),   
  some positive Fourier series, called   
 Virasoro characters encode critical exponents\cite{BPZ,LueM},  
Casimir coefficients  and
all multiplicities of physical states. 
At first sight they are defined on the upper half plane, 
but since they have fascinating modular properties, 
identifying the Riemann surface
on which they live and related symmetry groups is a 
progress. In this text we present some computations 
for $sl2$ WZW theories\cite{Wi}, but the general landscape 
is worth being observed:

After some years P. Bantay\cite{Ba1}  proved  W. Nahm's 
conjecture\cite{Nahm}:
representations 
(which we will generically call $ \rho $ in the sequel), 
of the modular group which occur 
in rational conformal theories, have principal congruence 
kernels, i.e. they are  representations of some 
$SL_2( \Z /N \Z ) $. 
 N is always the order of:  
$$ \rho (T)\ =\rho \left( \pmatrix{1 & 1\cr 0 &1 }\right) $$

As we show below, this theorem by Bantay brings a simplification in 
 identifying the action of the modular group. 
In practice, this means replacing  
an $SL_2( \Z ) $ matrix $ \pmatrix{ a & b\cr   c  & d} $, 
 by its residue matrix 
$ \pmatrix{  A & B\cr   C  & D} $  in  $SL_2( \Z / N\Z ) $.
Some nice formulae from "algebraic K-theory ", 
 coming from the existence of inverses in the ring 
 $  \Z / N\Z $, and various related lemmas 
 pave the way towards some compact formulae 
 within a bounded number of steps.  We therefore 
 somehow improve on L. Jeffrey's seminal 
computations\cite{Jef}.

 A consequence of this is
 also a better determination of the manifold of physical 
inequivalent $\tau '$s, a
 Riemann surface, which we propose to 
call {\bf "dynamical moduli space"}, because multiplicity of 
excited physical states ("descendants") 
is a truely  dynamical feature.

 When $N=q_1 \  q_2  \ldots \  q_r     $ , with each $q_i  $  being 
 a prime power, 
  decomposition of a  $ SL_2( \Z / N\Z ) $ group into its 
  $ SL_2( \Z /q \Z ) $ primary factors, (which means 
  restriction of the rep.  to the subgroups)
   can be performed as detailed in 
 a former text:
 for each factor, the generators are, for each prime $p_i $   
 dividing  $N$,   
\be T_i \ =  T^{c_i } , \quad  
    S_i \ =  T^{1-c_i }\  S T^{1-c_i }\ S T^{1-c_i } S^{-1}  \ee  
where we have used in the ring 
  $ \Z / N\Z  $ the canonical decomposition 
 into idempotents: $ 1  \equiv \ \sum c_i  \  $ mod $N$;    
 $ c_i  \equiv 1  \  $  mod $\     q_i        $ and 
 $      \equiv 0     $  mod the
 other   $  q_j      $'s.        
$S$ and  $T$ are the commuting products 
  of their primary images. 
This decomposition will be used in section $4$.  
\vskip 1.cm 

We consider the representation  $\rho $ of 
  dimension $n-1$ given by  
 \begin{eqnarray}
S_{\alpha \beta }\  := \rho 
 \left(   \pmatrix{0 & -1\cr 1 &0 }\right)_{\alpha \beta }=
\sqrt{{2\over n} } \  
  \ \sin \left( {\pi \alpha \beta \over n  }  \right) & &    \\   
T_{\alpha \beta }\    := \rho 
 \left(   \pmatrix{1 & 1\cr 0 &1 }\right)_{\alpha \beta } =
 \delta_{ \alpha \beta } \quad  
   e \left(  { \alpha^2   \over 4n } - { 1\over 8 } \right)&  &  \\   
 e( \tau )\  :=\exp{(2i\pi  \tau ) } \ , \  \alpha , \beta \in 
       \{1, \ldots , n-1    \}, \quad   & &    \nonumber  \\   
 \delta_{ \alpha \beta } =
\hbox{Kronecker's symbol with  values } 0, 1 
                                         &  &  \nonumber   
 \end{eqnarray}

In this representation $\rho $, N, the order of  $\rho (T) $, which 
we denoted  $n_{\infty }$
  in our   precedent texts,  is therefore 
 $4n $ if n is even, and $8n $ if n is odd. 

Our main concern in the following is 
therefore to  compute  images of 
 $2\times 2$  matrices using their decomposition as words 
in S and T mod N. 
 
\section{ $Gau\beta $ sums ingredients}

Since T is diagonal,  the technical point in computing 
 the image of a matrix, which is a word in 
 S and T, under representation $\rho $ above, is 
 being able to  have adequate expressions for the sum 
over $ \beta  $ appearing in: 
$$ \rho ( ST^C \  S )_{\alpha \gamma }  =\sum_{\beta } 
   S_{\alpha \beta }   \    T^C_{  \beta   \beta }  
 \     S_{ \beta   \gamma    }         $$

That is, the  key ingredient is 
 \begin{eqnarray}
{\cal T  }(\alpha  ,\gamma ,C ,n)\ &:= &
           \sum_{\beta =1}^{n-1} f(\beta )  \\  
\hbox{where }\        f(\beta )    &:= & 
          \sin \left( {\pi \alpha \beta \over n  }  \right) 
   \      \sin \left( {\pi \beta \gamma \over n  }  \right) 
   \       e \left( { C \beta^2       \over 4n }  \right)
 \end{eqnarray} 

Using the properties that:  
           $ f(\beta )=  f(-\beta )=  f(\beta +2n)$,
and  when  $n| \beta $ ,        $ \  f(\beta )=0  $ ;  
  \  we can as well 
express ${\cal T  } $ as a sum over  $\Z /n \Z $ ,
 $\Z /2n \Z $   , or    $\Z /4n \Z $. 

\vskip 1.cm
{\bf  When $ (C, 2n )=1 $} by completing the squares, we obtain: 

  \begin{equation}
{\cal T  }= {1 \over 8 }\ S(C,4n ) \ 
 \left[\ e \left( {-C^{-1 }( \alpha - \gamma)^2 \over 4n } \right) - 
         e \left( {-C^{-1 }( \alpha + \gamma)^2 \over 4n } \right) 
       \right]       
\end{equation}

where $C^{-1 } $ is any integer k  such that $ 4n| Ck-1 $ , in 
 view of having unified notations, we always take in the following 
 for   $C^{-1 } $ the inverse of C mod 8n, rather than mod 4n. 
 \begin{equation}
 S(C,N ) :=  \sum_{ \beta \in \Z /N\Z } \ 
   e  \left( { C \beta^2       \over N }  \right)
           \hbox{ is a }Gau\beta    \  sum. 
\end{equation}  
\vskip 1.cm  

{\bf  When $ C=\ tn  $} by splitting the sum between even and odd 
 $ \beta$'s, one obtains:
  \begin{equation}
{\cal T  }= {n \over 4 }\left[  
\left( 1+i^{C/n}\  (-1)^{ \alpha - \gamma \over n}  \right) 
                 \delta_{ \alpha ,\  \gamma \   mod n  }  - 
\left( 1+i^{C/n}\  (-1)^{ \alpha + \gamma \over n}  \right) 
                 \delta_{ \alpha ,\ - \gamma \  mod n  }  \right]       
\end{equation}  

{\bf  When $ C=\ 2\Gamma  \ , \  (\Gamma , n )=1 $} ,   
we have to distinguish according to the parity of  $  \alpha - \gamma $:
 if even, the square is obviously completed in the sum as above, whereas
 if it is equal to some  $2l+1$ , we need an extra lemma, making a sum 
  over $\Z /8n\Z $ enter the game:  
  \begin{eqnarray} 
{\cal T  } &=&    {i \over 2}\  
e   \left(  {  \Gamma^{-1} \   (\alpha^2 \  + \gamma^2 )\over 8 n
              }  \right) \   
sin     \left(  { \pi  \Gamma^{-1} \   \alpha \   \gamma \over 2 n
                   }\right) \    SF       \nonumber  \\  
  SF  &=& S(  \Gamma , 2n) \quad  \quad  \hbox{  for }  
  \alpha  -  \gamma   \hbox{ even } \\  
  SF  &=&   {1 \over 2}\    S( \Gamma , 8n) \  - S(  \Gamma , 2n) 
    \  \hbox{  for } \alpha -  \gamma   \hbox{  odd }
 \end{eqnarray}

\section{ Explicit compact expressions } 
\noindent 
\vskip .5cm 
{\bf  When $ (C, 2n )=1 $}   set   
   $U:= (A+1)C^{-1}  ,\  V  :=  (D+1)C^{-1}   $   
 , $ \zeta_8 = e(1/8 )  $.

 \begin{eqnarray}
 \rho \left( \pmatrix{  A & B\cr   C & D}
                          \right)_{\alpha  \lambda }
&= &   \rho \left( T^U \ ST^C \ ST^V\  
                          \right)_{\alpha  \lambda }  \\  
&= & { 2  \over n }  \  \zeta^{-2-U-V }_8  
{\cal T  }(\alpha  ,\gamma , C ,n)\  
  e \left(  { \alpha^2 \ U+ \lambda^2 \ V  \over 4 n }\right)  
     \nonumber  \\  
=  { 1 \over 2 n }  \  \zeta^{2-C-U-V }_8   \ S(C,4n)\ 
& &\sin \left( {\pi C^{-1} \ \alpha   \lambda \over n}\right)  
 e \left(  { C^{-1} \  
 (\alpha^2  A+ \lambda^2  D)\over 4 n }\right) 
 \end{eqnarray}  

Note that for $(C,2n)=1$,  $S(C,4n)$ is a Galois conjuguate of   
 the famous Dirichlet's $S(1,4n)=2\sqrt{n}\ (1+i)$, therefore is never zero; 
the matrix element for $ \alpha =1,  \   \lambda =2  $ 
 doesn't vanish either. Assuming known congruence results
for Jacobi forms , or Bantay's general theorem, we have therefore proven: 

\vskip 1.cm 
{\bf Proposition:} consider four integers 
satisfying $ad-bc=1$ ;   for the representation 
$\rho $ coming from affine Lie algebra $sl_2 $ at level 
 $k=n-2 $,   
 \begin{equation}
 \pmatrix{  a & b\cr   c & d} \in Ker\ \rho  \quad   \quad 
 \hbox{           implies that }
 (c,\ 2n) \neq 1  
  \end{equation}

\vskip .5cm 
\noindent 

{\bf  When $ (c, 2n )=1 $ and $n$ odd }, we can simplify the above
expression using Legendre's symbol (which takes here 
 only values $1$ or $-1$):    
\be  S(c,4n) = 2 (1+i^{nc})\  \left( {c  \over n}\right) 
                S(1,n)
\ee  
Since $mod\  8$, any odd residue satisfies, $ C\equiv C^{-1 }$, 
we can recast the eighth root of unity factors into the form
(which is a step towards a formula in next section):  

\begin{eqnarray} 
& & \rho \left( \pmatrix{  A & B\cr   C & D} 
         \right)_{\alpha  \lambda }
 =    \nonumber   \\ 
&=& \sqrt{ { 2  \over n }} \   \left( {C \over n}\right) 
  \  \zeta^{g(C,n)-(A+D+3)C }_8 \  
 \sin \left( {\pi C^{-1} \ \alpha   \lambda \over n }\right)  
  e \left(  { C^{-1} \  
 (\alpha^2  A+ \lambda^2  D)\over 4 n }\right) 
\end{eqnarray}
 where $g(C,n) $ is an integer which depends only on 
the residues of $c$ and $n$ $mod\  4$: 
\bea 
\hbox{if }\  \  \  \    C\equiv  1 \ mod \ 4,\ 
                             & g=3 & \nonumber \\  
\hbox{if }  C\equiv -1 \     
\hbox{and } n\equiv  1 \ mod\ 4,\  &g=1 &\nonumber \\  
\hbox{if }  C\equiv -1 \ 
\hbox{and } n\equiv -1 \ mod\ 4,\  &g=-3&           
\eea  

A first check of this formula is that it gives the same image for 
a matrix and its opposite, this indeed results from properties 
of Legendre's symbol and from: 
\be g(C, n )  -g(-C, n )+2C\ \equiv \ 2(n+1) \  mod\  8   
\ee 
\vskip 1.cm 
{\bf  When $ (d, 2n )=1 $}, denoting  as above by $D^{-1}$ the inverse mod 
 $ 8n $ and $X:= - (C+1)D^{-1}$, $Y := (B-1)D^{-1}$ we get:  
$$
 \rho  \left( 
 \pmatrix{A & B\cr C & D} \right)_{\alpha ,\lambda }
=  \rho (  T^X\ ST^{-D}\ ST^Y S)_{ \alpha ,\lambda }      $$

%
 \begin{equation}
=\left( {2\over n} \right)^{3/2 } {1\over 4}   
 \    \zeta^{D-X-Y-2}_8 
S(-D, 4n ) e\left( {  \alpha^2 BD^{-1} \over 4n} \right)  \  
{\cal T }( \alpha D^{-1} ,  \lambda , -CD^{-1}  ,n )    
 \end{equation}

This gives in particular, using above results ( $\delta $ symbols 
 mod n can be recast into equalities mod 2n ), a decorated Galois 
permutation (notice $A\equiv D^{-1}$ mod N):      
 \begin{eqnarray} 
 \rho  \left( 
    \pmatrix{  A & B\cr   0 & D}\right)_{ \alpha ,\lambda }
 &=&  {1\over \sqrt{8n} } \  \zeta^{D-X-Y-2}_8 \ S(-D, 4n )  \ 
    e\left( { AB \alpha^2 \over 4n} \right)  \nonumber \\   
 .& & \left[ \delta_{\alpha D^{-1}, \ \lambda \ mod\  2n} -
        \delta_{\alpha D^{-1},\ -\lambda \ mod\ 2n} \right]
 \end{eqnarray}

 \begin{equation}
= \pm \zeta^{2(A-1)-AB}_8 e\left( { AB \alpha^2 \over 4n}\right) 
 \left[ \delta_{A\alpha ,\  \lambda \ mod\ 2n} -
        \delta_{A\alpha ,\ -\lambda \ mod\ 2n} \right]
 \end{equation} 
where the sign is studied in \cite{CDL}.
\smallskip

\section{ Galois properties }
C. Itzykson and J. Lascoux early recognized the relevance of 
classical Galois theory for study of CFT's. Here let us comment 
on the "etat de l'art", using the simplicity of 
 the matrices involved, allowing short trigonometric 
expressions:    
  
When d and n are $>0 $ , and $(d,2n)=1 $ , quadratic reciprocity law 
applied to odd factors of n  gives:   
\begin{eqnarray} 
\sigma_d \  ( S_{ \alpha ,\beta } ) 
 &=& \left( { -2n\over d} \right)  
 \sqrt{{ 2 \over n} } \  
 \sin \left( {\pi d   \alpha \beta \over n}\right) \\  
 &=& \left( { -2n\over d} \right)
 \ sign( n - <  \alpha d >_{2n} )\ 
 S_{\sigma_d     (  \alpha ) ,\beta } 
 \end{eqnarray} 
where $<u >_{2n} $ is the number $ \in \{0,\dots ,2n-1  \}   $ congruent
to $u$ mod  $2n$.  Here we do not suppose $n$ odd, notice that 
 if an odd power of $2$ enters into the decomposition of  $n$, 
$ \left( { 4\over d} \right)  =1$ and there is no $ \sqrt{2}  $ in the
formula.   
\begin{eqnarray} 
\hbox{if }  \alpha d= 2ln+  \gamma \ \ \   
  \sigma_d \  (  \alpha  )   &:=& \gamma     \nonumber \\ 
\hbox{whereas if }   \alpha d= (2l+1)n+  \gamma \ \ \  
   \sigma_d \  (  \alpha  )  &:=&   n - \gamma     
 \end{eqnarray} 
  
{\bf Proposition: } When $L$ is invertible in $\Z/8n \Z$, 
the cyclotomic action on the image of {\bf any }
 matrix is\footnote{a.c. thanks M. Bauer for, 
a long time ago, pointing this}: 
\be  
\sigma_L  \left(  \rho  \left( 
 \pmatrix{A & B\cr C & D} \right) \right) = 
 \rho  \left( \pmatrix{A & BL\cr CL^{-1} & D} \right)
\ee 
Proof: Since left and right hand sides are group morphisms 
and that $SL_2 $ is generated by $T$ and $S$ , it suffices 
to check it for $T$, this is obvious, and for $S$:
 Look therefore at the explicit expressions
 obtained for the image of a 
matrix with $C=1$, or even $C$ invertible,
 distinguishing again  the cases,
 we see another little miracle for $n$ odd:
\be  \left( {C \over n}\right) 
  \  \zeta^{g(C,n)\ -3C }_8 \        =   
   \left( {-2n \over C^{-1} }\right) = 
   \left( {-2n \over C      }\right) = 
(-1)^{{c^2 -1\over 8}+{c-1\over 2}{n+1\over 2}  }
 \  \left( {C \over n}\right) 
\ee 
Which is just what is needed to insure equality of the 
directly computed matrix and the Galois image of 
 $\rho (S) $.

Finally for any invertible  $C$ mod $8n$ :
\bea 
 \rho  \left( 
\pmatrix{A & B\cr C & D} \right)_{ \alpha \lambda } 
 &=&  
\sigma_{C^{-1}}  \left( \rho  
        \left(   \pmatrix{A & BC\cr 1 & D}
 \right)_{ \alpha \lambda }   \right)  =    \\ 
 \rho ( T^{(A+1)C^{-1} } \ ST^{C}\ ST^{(D+1)C^{-1} }
 \ )_{ \alpha \lambda }
 &=& \sigma_{C^{-1}}
(\rho ( T^{A}\ ST^{D})_{ \alpha \lambda } ) \\   
  = \left( { -2n\over C} \right)  
 & \sqrt{{ 2 \over n} }     & \  
 \sin \left( {\pi C^{-1}  \alpha \lambda \over n}
  \right) \          e \left( 
 {C^{-1}\  (A \alpha^2 +D\lambda^2 )\over 4n} \right) 
\eea

In \cite{Ba1} P. Bantay gave a 
 very interesting criterion for a matrix
 $ \pmatrix{  a & b\cr   c & d} $ with $(d,2n)=1$, 
to be in the kernel. It is related to generation properties 
of finite unimodular matrix groups \cite{Behr, Harpe}.   
 Although we will 
 have much more to say in later studies,  
we can confirm that the above 
 formula inserted in that criterion 
and the direct determination 
  of the kernel for some values 
of n  agree with our direct investigations which
we begin here to present  in  section 6. 
             
  Explicitly, $  \sigma_d \  (S) T^b =T^c\ S $ reads: 
\be  \left( { -2n\over d} \right) 
\sin \left( {\pi d\alpha \beta   \over n} \right)
e \left( {\beta^2 b - \alpha^2 c \over 4n}\right) 
  \zeta_8^{c-b} = 
      \sin \left( {\pi \alpha \beta   \over n} \right)
\ee   
{\bf for all }$  \alpha ,\beta =1,\dots ,n-1$. 
Taking the norm 
gives 
$$ \sin \left({\pi \alpha \beta (d-1)\over n}\right)
 \ \sin \left({\pi \alpha \beta (d+1)\over n}\right)=0
$$ 
This implies that if $(d,2n)=1 $, $d=nL +d_0 $ with 
$d_0 =\pm 1 $. Then   
$$ \sin \left( {\pi d\alpha \beta \over n}\right)
 = d_0  (-1)^{L\alpha \beta } 
 \sin \left( {\pi \alpha \beta \over n} \right)
$$
Therefore, for $n$ bigger than 4, $L=2l $, 
  and we distinguish:
   
{\bf For $n$ odd}, above eqs. for all 
$ \alpha ,\beta   $ require a sign equality 
\be  
b=4nb' \ \ , c=4nc'  \ \  ,  (-1)^{b'-c'}=
 \varepsilon (d\ ,n)
\ee 
where one can, distinguishing the values of $d_0 $, 
reshuffle this into: 
\be
 (-1)^{b'-c'}=
  \varepsilon (2ln+d_0 \ ,n)= 
   (-1)^{{l\ (\ l\ -d_0 ) \over 2}}
\ee 
We therefore obtain the following 
expression for the Bantay's criterion: 
 
The matrices $ \pmatrix{ a & b\cr c &d} $ which 
have $(d,2n)=1$, and are in the kernel, are for 
$n$ odd, those congruent 
mod $8n$ to :
  
\bea 
&    \pmatrix{  1  & 0 \cr  0 & 1}& ,\ 
     \pmatrix{  1  & 4n\cr 4n & 1} ,\ 
     \pmatrix{2n+1 & 0  \cr 0 & 2n+1} ,\ 
     \pmatrix{2n+1 & 4n \cr 4n& 2n+1} ,\ 
 \nonumber \\  
&    \pmatrix{2n-1 & 4n \cr 0 & 2n-1}& ,\ 
     \pmatrix{2n-1 & 0  \cr 4n& 2n-1} ,\ 
     \pmatrix{4n+1 & 0  \cr 4n& 4n+1} ,\ 
     \pmatrix{4n+1 & 4n \cr 0 & 4n+1} ,\ 
 \nonumber 
\eea 
and their opposite.

{\bf For $n$ even} 
we find that the same criterion picks matrices congruent
{\bf  mod $4n$ } to   
$$
  \pmatrix{  1  & 0 \cr  0 &    1} ,\ 
  \pmatrix{2n+1 & 0 \cr 0  & 2n+1} ,\ 
$$
and their opposite. We will comment elsewhere,  
in collaboration with P. Bantay and friends, we hope, 
 on this criterion, which applies to conjugacy 
classes, and its 
relationship with the enumeration of 
 $   SL_2( \Z /N \Z ) $ we have sketched.

\section{ A general  theorem }
 In contradistinction with  the first sections, 
where the computations where self contained 
we now rely fully  on established general results, in particular 
  the theorem that the modular representation is defined 
 in a cyclotomic field $\Q (\zeta_M )$. 
It has been proven\cite{Ba1} that   $M$ equals 
the conductor $N$. It would anyway be no trouble to 
enlarge the field if needed for further purpose. We therefore 
 have at disposal the Galois morphisms which we denote as 
always by $\sigma_{ L } $ . If we were reasoning with 
 $c \in \Z $, with   $ ch - qM =1 $, we could use the absolute 
Galois morphism 
   $\sigma_{ h  } = \sigma_{C^{-1} } $ here. The way we 
successfully tackled the eighth roots of unity, going from 
$n$ to $8n$ when needed  
in the first sections illustrate this.  

To make the notations  
lighter, we drop in this section the reference to 
$\rho $  , calling $S$ and $T$ the images by $\rho $ 
of the   $SL_2 (\Z )$ matrices defined in the beginning,
 as Bantay does. 
\vskip .5cm   
 
{\bf Theorem 1: }   In any RCFT, if $N$ is the order of $T$, 
and $C$ invertible mod $N$, 
\bea 
\rho \left( \pmatrix{ A & B \cr C & D} \right) &=& 
\sigma_{C^{-1} } (\ T^A \ ST^D \ ) \\  
\hbox{where } \ \ 
\sigma_{C^{-1} }  &\in & Gal(\Q (\zeta_N )/ \Q ) \nonumber
\eea   
{\bf Proof: } 
          
$ \sigma_{C^{-1} } (T^A \ ST^D )
 =T^{AC^{-1}}\  \sigma_{C^{-1} } (S)\ T^{DC^{-1}} $.
 
But relying on theorems proven by T. Gannon, J. Lascoux 
and the author, Bantay has established \cite{Ba1} 
\footnote{Of course many people contributed significantly 
to the development of this field: 
J. de Boer, C. Goeree, C. Itzykson, W. Nahm, J.B. Zuber, 
V. Pasquier, P. Ruelle, E. Thiran, J. Wyers, D. Altschuler, 
 M. Bauer, \dots  } that: 
\be 
 \sigma_{C^{-1} } (S)\ = T^{ C^{-1}}\  ST^C \ ST^{ C^{-1}}
\ee 
Therefore 
$$ 
 \sigma_{C^{-1} }  (T^A  ST^D )= 
 T^{AC^{-1} +C^{-1}}S T^C S 
 T^{ C^{-1} +DC^{-1} }       \ = 
\rho \left( \pmatrix{ A & B \cr C & D} \right)
$$ 

{\bf Theorem 2: }  
  
In any RCFT, where $N$, the order of
 $\rho \pmatrix{ 1 & 1 \cr 0 & 1 } $, 
 doesn't divide $12$,

    {\bf If }
 $\pmatrix{ a & b \cr c & d} \in Ker \  \rho $,  
{\bf Then } $(c,N)\neq 1 $. 
    
{\bf Proof : }If  $(c,N)= 1 $, one would have 
 $\sigma_{C^{-1} }  (T^a  ST^d )=  Id $, therefore 
  $S^{-1}\ = T^{a+d} $.Thus S and T commute, which implies 
 $  S^{-1}\ = T^3 $. 
  
Therefore $N$ divides $a+d-3 $, $4(a+d)$,  and $3(a+d+1)$.
               
If  $a+d $ is even , N is odd since it divides $a+d-3 $, 
but since it divides  $4(a+d)$, it should divides both 
 $a+d$ and $3$.
 
If  $a+d=3 $, N divides $12$. 
If  $a+d\neq 3 $ is odd , N is even 
 since it divides  $a+d-3$,  
but then  it also divides $a+d+1$, thus it 
divides    $4$.

\section{ A genus formula }
{\bf Proposition: } when $n\equiv 3 \ mod \ 4 $ 
 characters of affine $sl2$ algebra at level $k=n-2$, 
bring a representation of $   SL_2( \Z /8n \Z ) $ 
which is separately a representation of
 $ SL_2( \Z /8 \Z )/\pm 1 $ and of 
 $ SL_2( \Z /n \Z )/\pm 1 $ . The kernel 
of the first one is exactly 
\be \{  \pmatrix{  1 & 0\cr   0 & 1} ,\ 
        \pmatrix{  1 & 4\cr   4 & 1} ,\ 
        \pmatrix{  5 & 4\cr   0 & 5} ,\ 
        \pmatrix{  5 & 0\cr   4 & 5}  \ 
   \} \hbox{ in }\   SL_2( \Z /8 \Z )/\pm 1 
\ee  
Proof: when n is odd, $ 1= c_2 +c_n $ , with $  c_2 = n^2 $ , 
$c_n =1-n^2  $ is a decomposition into idempotents, because any 
 odd number is a square root of 1 mod 8. The restriction of 
the representation to the   $ SL_2( \Z /8 \Z ) $ 
subgroup is therefore given by images of 
  $S_2 $ and $ T_2=T^{n^2} $ as explained in section 1. 
\be 
\rho (T_2 )_{\alpha , \lambda }\ =
     \delta_{\alpha ,\lambda }\ 
  e\left( { n \alpha^2 \over 4} - {1\over 8 }\right)
\ee  
\be 
 \hbox{implies that }\ \  \rho (T_2 )^4 \  = 
\ - Id_{ \C^{n-1}} \ \hbox{is central} 
\ee 
\be 
 \hbox{since  }\ \ 
S_2   \equiv \pmatrix{  1-n^2  &-n^2\cr n^2 & 1- n^2 }\ , \  
S_2^2 \equiv \pmatrix{  1-2n^2 & 0  \cr   0 & 1-2n^2 }\ 
           \in  SL_2( \Z /8n \Z )\ , 
\ee 
for $n\equiv 3 \ mod\ 4$ 
 the formulae of section 3 above give 
$ \rho ( S_2^2 )= Id $. This with centrality of      
   $ \rho (T_2 )^4 $, is sufficient to prove 
that the four above matrices (which are words in 
$ S_2 $ , $T_2 $) are in the kernel. We have already 
 discussed in detail the conjugacy classes structure 
of   $ SL_2( \Z /8 \Z ) /\pm 1$, a group with $192$ 
elements, in a previous work \cite{Cinv}.
 Computer packages  
can also give a lot of useful outputs for cross-checks.  
Here the cardinal of the image of this group 
should divides $192/4 = 48 $ , and should be a multiple 
 of $8$, the order of $\rho (T_2 )$. This requirement 
 implies that the kernel can only be the normal 
subgroup given by the four matrices given above.  
  
Let us comment an example: applying Bantay's 
criterion we find that  for $n$ odd
$$ \pmatrix{  2n+1  &4n \cr 4n &  2n+1 }\ \in 
  Ker \rho \ \bigcap \  SL_2( \Z /8n \Z )       $$ 
This matrix is
 $ \equiv  \pmatrix{  2n+1  &4 \cr 4 &  2n+1 } $ mod $8$. 
For $n\equiv 3 $ mod $4$, which is the case of our prop. 
above, it is 
 $ \equiv  \pmatrix{-1  &4 \cr 4 &-1 } $
which is one of the element of the kernel we identified in 
  $SL_2( \Z /8 \Z )/\pm 1$.

{\bf Proposition: } for $n=p $  a prime bigger or equal 
 to $7$ and congruent to $7$ mod $4$ , the genus 
 of the completed 
 (smooth, compact, without punctures ) Riemann surface, on which 
characters live,  is:    
\bea 
g &=&\  1 + 12 \ p(p^2\ -1)\ 
\left(  {1 \over 6 }- {1 \over 8p } \right) \nonumber \\  
 =\  1+ { (p^2\ -1)(4p-3) \over 2 }\  
  &=&\ { 4p^3\ -3p^2 \ -4p+5 \over 2 }
\eea
To make the proof obvious let us first notice:

{\bf lemma:} Let $\rho $ be a representation of 
$G=G_1 \times G_2 $,  a direct product, denote the projection 
by: 
\bea   Ker \  \rho \ &\longrightarrow &
 \ G_1 \times G_2                       \nonumber \\  
 g \ &\longrightarrow & ( \varphi_1(g),\ 
            \varphi_2(g) )= (g_1, \ g_2 ) 
\nonumber 
\eea    
Then $ \varphi_2(Ker \rho ) =\{g_2,\hbox{ for which exists }
   g_1,  \rho (g_1 g_2 )=1  \}$ is a normal subgroup of 
 $G_2 $ . This is trivial: 
$$  \rho ( hg_1 h^{-1}\  hg_2 h^{-1})\  = 1 $$ 
Now,  excepting primes 2 and 3, $ SL_2( \Z /p \Z )/\pm 1 $ is simple 
 and of order $p(p^2\ -1)/2 $, Therefore we showed above that 
when $n=p\equiv 7 \ mod 4 $, the image of the kernel in 
 $ SL_2( \Z /p \Z ) $ is exactly $ \pm Id $.
 therefore $Im\  \rho $ is a group 
 with exactly $   48 \ \times  {p (p^2\ -1) \over 2 }$ 
elements. Expression of the genus is given by Riemann-Hurwitz 
 formula for a triangulation with $8p$-valent vertices. For 
 $p=7$, $g= 601$, which is big, but $g-1$ is 
already eight times smaller 
  than the corresponding value  
for principal curve  $X(8p) $.  When $n$ factors into many 
 powers, each $\rho (S_i )^2 $ is a central element of the
 representation, not necessarily a constant.

\section{Appendix}
Here we give some explicit 
expressions of characters for the interested 
reader. First we have the  useful expansions: 

$$ {1\over \prod (1-q^n )^3 } = 1+3q +9q^2 +22 q^3 +51 q^4 
   +108 q^5  +221 q^6 +429 q^7 +810 q^8 +\dots     $$
\bea -ln (  \prod (1-q^n )\ ) &=& \ \sum_{k> 0 } 
  {\sigma_1 (k)\ q^k \over k }    \nonumber \\  
&=& {q\over 1-q}\ +  \sum_{p \ prime  } {q^p \over p} 
 \  +{3 \over 4 }q^4 +q^6+
     {7 \over 8 }q^8 +\dots  \nonumber 
\eea 
  $$ \hbox{ where } \ 
 \sigma_1 (k= \prod p^{\nu } ) = 
      \sum_{m|k} \ m =
 \prod { (p^{\nu +1}\ -1 ) \over (p-1) } $$

The formula for $sl_2 $ characters 
 labelled by   shifted weights 
$ \lambda = 1,\dots ,n-1 $ is\cite{Ka1,ID}: 
 
\be 
\chi_{\lambda ,[n] } \  =  {1\over \eta^3}\  
\sum_{x\equiv \lambda \ mod\ 2n} \ x\ 
 q^{{x^2 \over 4n}}
\ee 
We have been concerned with the representation 
 underlying the identity: 
\be 
   \chi_{\alpha } \left( {-1\over \tau }\right)  \  =  
\sum_{\beta =1 }^{n-1}\ \rho (S)_{\alpha ,\beta }
 \ \chi_{\beta  }(\tau ) 
\ee 
Let us focus on the case $n=3$  
(i. e. $k=1$),  the two characters are:  
\bea 
\chi_1 &=& q^{{-1\over 24}} \  
{\sum_{l\in \Z }\ (1+6l)\ q^{l(1+3l)} \over 
  \prod (1-q^n )^3  }  \nonumber \\  
       &=& q^{{-1\over 24}} \  (1+3q+4q^2 +7q^3 + 13q^4 +
  19q^5 + 29q^6 +43q^7 +62q^8 +90q^9 +\dots ) \\  
\chi_2 &=& 2\  q^{{5\over 24}} \  
{\sum_{l\in \Z }\ (1+3l)\ q^{l(2+3l)} \over 
  \prod (1-q^n )^3  }  \nonumber \\ 
       &=&   q^{{5\over 24}} \ (2+2q+6q^2 +8q^3 + 14q^4 +
  20q^5 + 34q^6 +46q^7 +70q^8 +96q^9 +\dots )  
\eea  
They satisfy  
\be 
\chi_1  \chi_2 \ (\chi_1^4 \ -\chi_2^4 )\ = 2  
\ee   
 
Therefore our "dynamical moduli space" is here, when we adopt 
the $\chi $'s as coordinates, a smooth complex curve, 
of genus $10$, degree $6$, 
here embedded in projective space $\P^2 $. 
It has six points at infinity.

Since the isometry group, which is a Galois group for an extension 
of the function field $\Q (j) $ is here solvable,  we can solve 
the above equation by radicals, which means  we can parametrize 
our "dynamical moduli space" in terms 
 of a single complex number $t$: 
\be 
\hbox{set  }t := \chi_1  \chi_2 \ \   \hbox{then  }
 \chi_1^8 -{2\over t} \chi_1^4 \ -t^4=\ 0 
\ee  
\bea 
 \chi_1 &=&  i^{ \alpha }\  t^{-1\over 4}\ 
 (1\pm \sqrt{1+t^6} )^{ 1\over 4}  \nonumber \\ 
 \chi_2 &=&  i^{-\alpha }\  t^{5\over 4}\ 
 (1\pm \sqrt{1+t^6} )^{-1\over 4}  
\eea 
The already high genus, is due to the 
cuts needed in taking roots. This case is exceptional 
in the sense that for higher $n$'s the  $j(\tau ) $ 
function enters the game.

\bigskip\noindent{\bf  Acknowledgements:}
 
Without kind personal support of J. Wolfart, this work 
 wouldn't have been possible. We'd like also to thank
for 
hospitality or nice scientific conversations:
 H. Behr, M. Pflaum und  die Fachbereich Math. 
und Rechnenbetrieb Leute   
   der   Goethe  Univ. , M. L\"uscher, R. Stora, 
 P. de la Harpe, A. Alekseev,J. Gasser, 
 P. Minkowski, einige Fachleute aus Z\"urich und Ihes, 
 J.P. Derendinger, D. Altschuler, J. Lascoux,
M. Giusti, M. Streit,  G. Kemper, E. Cremmer, B. Doucot, 
J. Magnen, P. Viot, V.Pasquier.

\end{document}